\documentclass{elsart5p}

\usepackage{natbib}

\usepackage{graphicx}

\usepackage{amssymb}

\def\simless{\mathbin{\lower 3pt\hbox
   {$\rlap{\raise 5pt\hbox{$\char'074$}}\mathchar''7218$}}}   
\def\simgreat{\mathbin{\lower 3pt\hbox
   {$\rlap{\raise 5pt\hbox{$\char'076$}}\mathchar''7218$}}}   

\newcommand{\be}{\begin{equation}}
\newcommand{\ba}{\begin{eqnarray}}
\newcommand{\ee}{\end{equation}}
\newcommand{\ea}{\end{eqnarray}}
\newcommand{\url}{\tt}%
\def\lesssim{\mathrel{\hbox{\rlap{\hbox{\lower4pt\hbox{$\sim$}}}\hbox{$<$}}}}
\def\gtrsim{\mathrel{\hbox{\rlap{\hbox{\lower4pt\hbox{$\sim$}}}\hbox{$>$}}}}

\begin{document}

\begin{frontmatter}



\title{The Kinetic Sunyaev-Zel'dovich Effect from Patchy Reionization: the View
  from the Simulations }


\author{Ilian T. Iliev$^1$, Ue-Li Pen$^1$,
J. Richard Bond$^1$, Garrelt Mellema$^{2}$, }
\author{Paul R. Shapiro$^3$}

\address{$^1$Canadian Institute for Theoretical Astrophysics, University
  of Toronto, 60 St. George Street, Toronto, ON M5S 3H8, Canada\\
$^2$Stockholm Observatory,
AlbaNova University Center,
Stockholm University,
SE-106 91 Stockholm,
Sweden\\
$^3$Department of Astronomy, 1 University Station, C1400,
  Austin, TX 78712, USA\\
E-mails:  iliev@cita.utoronto.ca, pen@cita.utoronto.ca, bond@cita.utoronto.ca,
garrelt@astro.su.se,  shapiro@astro.as.utexas.edu}

\begin{abstract}
We present the first calculation of the kinetic Sunyaev-Zel'dovich (kSZ) effect 
due to the inhomogeneus reionization of the universe based on detailed 
large-scale radiative transfer simulations of reionization. The resulting sky
power spectra peak at $\ell=2000-8000$ with maximum values of 
$[\ell(\ell+1)C_\ell/(2\pi)]_{\rm max}\sim4-7\times10^{-13}$. The scale roughly
corresponds to the typical ionized bubble sizes observed in our simulations, 
of $\sim5-20$~Mpc. The kSZ anisotropy signal from reionization dominates the 
primary CMB signal above $\ell=3000$. At large 
scales the patchy kSZ signal depends only on the source efficiencies. It is
higher when sources are more efficient at producing ionizing photons, since
such sources produce larger ionized regions, on average, than less efficient
sources. The introduction of sub-grid gas clumping in the radiative transfer 
simulations produce significantly more power at small scales, but has little
effect at large scales. The patchy reionization kSZ signal is dominated by the
post-reionization signal from fully-ionized gas, but the two contributions
are of similar order at scales $\ell\sim3000-10^4$, indicating that the kSZ 
anisotropies from reionization are an important component of the total kSZ 
signal at these scales.
\end{abstract}


\end{frontmatter}

\section{Introduction}
\label{intro_sect}

The secondary anisotropies of the Cosmic Microwave Background (CMB)
are emerging as one of the most powerful tools in cosmology. The
small-scale anisotropies are probes of cosmological structures and are
thus a valuable tool for studying their formation and properties. The
key effect resulting in small-scale CMB anisotropies is the
Sunyaev-Zel'dovich (SZ) effect \citep{1969Ap&SS...4..301Z}, produced
by Compton scattering of the CMB photons on moving free electrons.
When this is due to thermal motions the effect is referred to as
thermal SZ effect (tSZ), while when it is due to the electrons moving
with a net bulk peculiar velocity it is called kinetic SZ effect (kSZ)
\citep{1980MNRAS.190..413S}. The former, coming largely from hot gas
in low-redshift galaxy clusters, is the dominant effect, but has a
different spectrum than the CMB primary anisotropies, with a
characteristic zero at $\sim217$~MHz, so its contribution can be
separated. By contrast, kSZ anisotropies have a spectrum identical
with that of the primary anisotropies. Separation of its effects are
possible only by virtue of its differing spatial structure. We
consider the kSZ fluctuations as composed of two basic contributions,
one coming from inhomogeneous reionization and the other from the
fully-ionized gas after reionization. The calculation of the patchy
reionization contribution to the kSZ signal based on the first
large-scale radiative transfer simulations of reionization is the main
focus of this paper.  This signal provides a signature of the
character of reionization that will complement other approaches like
observations of the redshifted 21-cm line of hydrogen and surveys of
high-z Ly-$\alpha$ emitters
\citep[e.g.][]{2000ApJ...528..597T,2002ApJ...572L.123I,2003AJ....125.1006R,
2004ApJ...604L..13S,2004ApJ...608L..77G,2004MNRAS.347..187F,
21cmreionpaper,2005astro.ph.11196M,2005astro.ph.12516S,2006NewAR..50...94B}.

The kSZ from a fully-ionized medium has been studied previously by
both analytical and numerical means. When linear perturbation theory is 
used, the effect is associated with the quadratic nonlinearities in the 
electron density current and is usually referred to as the Ostriker-Vishniac 
effect \citep{{1986ApJ...306L..51O},{1987ApJ...322..597V}}. Calculating the
full, nonlinear effect is much more difficult to do analytically,
although some models have been proposed
\citep[e.g.][]{1998PhRvD..58d3001J,2000ApJ...529...12H,
2002PhRvL..88u1301M,2004MNRAS.347.1224Z}.  This effect involves
coupling of very large-scale to quite small-scale density fluctuation
modes and thus requires a very large dynamic range in order to be simulated
correctly. Current simulations are in rough agreement, but have not
quite converged yet
\citep[e.g.][]{2001ApJ...549..681S,2002PhRvL..88u1301M,2004MNRAS.347.1224Z}.
All simulations predict significant enhancement of the small-scale
anisotropies compared to the analytical theory, due to the nonlinear
evolution.

Calculations of the second important  kSZ contribution, from patchy
reionization, are even more complicated. In addition to the above
difficulties, it also requires detailed modelling of the radiative transfer to
derive the sizes and distributions of H~II regions in space and time (the
reionization geometry) and how these correlate with the velocity and density 
fields. To date this problem has been studied by only a few recent works. Most 
of these estimates were done by semi-analytical models, which used different 
simplified approaches to model the inhomogeneous reionization 
\citep{1998ApJ...508..435G,
2003ApJ...598..756S, 2005ApJ...630..657Z,2005ApJ...630..643M}. While such 
models are useful since they are simple and much cheaper to calculate 
than full radiative transfer simulations, and are thus convenient for
exploring various scenarios, their results must be checked against full 
detailed simulations in order to ascertain their reliability. The only two 
existing numerical studies of this effect, \citet{2001ApJ...551....3G} and 
\citet{2005MNRAS.360.1063S} used fairly small computational boxes 
($4\,\rm h^{-1} Mpc$ and $20\,\rm h^{-1} Mpc$, respectively), and, as a
consequence, the results they derived significantly underestimate the signal 
since such small boxes do not include the larger-scale velocity modes which 
are quite important.  

There are several upcoming experiments which would search for kSZ effect, in
particular the Atacama Cosmology Telescope 
(ACT)\footnote{$\url{http://www.hep.upenn.edu/\!\sim\!angelica/act/act.html}$} 
and South Pole Telescope 
(SPT)\footnote{$\url{http://spt.uchicago.edu/science/index.html}$}, both of 
which are expected to be operational by early 2007. These experiments would 
have $\sim1'$ resolution and $\sim\mu$K sensitivities, which should be 
sufficient to detect the kSZ effect.  

In this paper we present the first calculations of the kSZ effect from patchy
reionization based on large-scale radiative transfer simulations. We simulate
reionization using a $(100\,\rm h^{-1}Mpc)^3$ simulation volume, which is
sufficient to capture the relevant large-scale density and velocity
perturbations, an important improvement over previous efforts. We use the 
ray-tracing code $C^2-Ray$ \citep{methodpaper} to follow the radiation from 
all ionizing sources in that volume identified with the resolved halos, which 
are of dwarf galaxy size or larger. The halos and underlying density field are 
provided by a very large N-body simulation with the code PMFAST 
\citep{2005NewA...10..393M}. Our methodology, simulation parameters and
results on the reionization character, geometry and observability of the 
redshifted 21-cm line of hydrogen were discussed in detail in
\citet[][Paper I]{2006MNRAS.369.1625I} and \citet[][Paper II]{21cmreionpaper}.
 
\section{Simulations}
\label{simul_sect}


Our simulations follow the evolution of a comoving simulation volume
of $(100\,h^{-1}\rm Mpc)^3$, corresponding to an angular size
$\sim1$~deg.  Our radiative transfer grids are of sizes $203^3$ or
$406^3$, which allows us to derive reliably the angular sky power
spectra for $l\approx 430-90,000\, (180,000)$. We utilize the flat-sky
approximation, which is appropriate for such relatively small angular
sizes. Our simulation methodology and parameters were described in
detail in Papers I and II. 
Here we provide just a brief summary. We start with performing a very 
large pure dark matter simulation of early structure formation, with 
$1624^3$ particles and $3248^3$ grid cells using the code PMFAST
\citep{2005NewA...10..393M}. This allows us to reliably identify (with
100 particles or more per halo) all halos with masses
$2.5\times10^9M_\odot$ or larger. We find and save the halo
catalogues, which contain the halo positions, masses and detailed
properties, in up to 100 time slices starting from high redshift
($z\sim30$) to the observed end of reionization at $z\sim6$. We also
save the corresponding density and bulk peculiar velocity fields at
the resolution of the radiative transfer grid. Unfortunately,
radiative transfer simulations at the full grid size of our N-body
computations are impractical on current computer hardware.

All identified halos are assumed to be sources of ionizing radiation and 
each is assigned a photon emissivity proportional to its mass, $M$, according 
to
\be 
  \dot{N}_\gamma=f_\gamma\frac{M\Omega_b}{\mu m_p t_s\Omega_0},
\ee
where $t_s$ is the source lifetime, $m_p$ is the proton mass, and $f_\gamma$ 
is a photon production efficiency which includes the number of photons 
produced per stellar atom, the star formation efficiency (i.e. what fraction
of the baryons are converted into stars) and the escape fraction (i.e. how
many of the produced ionizing photons escape the halos and are available to
ionize the IGM).

The radiative transfer is followed using our fast and accurate ray-tracing
photoionization and non-equilibrium chemistry code $C^2-Ray$. The code has been 
tested in detail for correctness and accuracy against available analytical
solutions and a number of other cosmological radiative transfer codes
\citep{methodpaper,2006astro.ph..3199I}. The radiation is traced from every 
source on the grid to every cell.
 
We have performed four radiative transfer simulations. These share the source 
lists and density fields given by the underlying N-body simulation, but adopt 
different assumptions about the source efficiencies and the sub-grid density 
fluctuations. The runs and notation are the same as in Paper II: runs f2000 
and f250 assume $f_\gamma=2000$ and 250, respectively, and no sub-grid gas 
clumping, while f2000C and f250C adopt the same respective efficiencies, 
$f_\gamma=2000$ and 250, but also add a sub-grid gas clumping, 
$C(z)=\langle n^2\rangle/\langle n\rangle^2$, which evolves with redshift 
according to
\be
C_{\rm subgrid}(z)=27.466 e^{-0.114z+0.001328\,z^2}.
\label{clumpfact_fit}
\ee
The last fit was obtained from another high-resolution PMFAST N-body 
simulation, with box size $(3.5\,\rm h^{-1}~Mpc)^3$ and a computational 
mesh and number of particles of $3248^3$ and $1624^3$, respectively. These 
parameters correspond to a particle mass of $10^3M_\odot$ and minimum 
resolved halo mass of $10^5M_\odot$. This box size was chosen so as to 
resolve the scales most relevant to the gas clumping - on scales smaller than
these the gas fluctuations would be below the Jeans scale, 
while on larger scales the density fluctuations are already present in our 
computational density fields and should not be doubly-counted. The expression
in equation~(\ref{clumpfact_fit}) excludes the matter inside collapsed 
minihalos (halos which are too small to cool atomicly, and thus have
inefficient star formation) since these are shielded, unlike the generally 
optically-thin IGM. This self-shielding results in a lower contribution of 
the minihalos to the total number of recombinations than one would infer 
from a simple gas clumping argument
\citep{2004MNRAS.348..753S,2005MNRAS...361..405I}. The effect of minihalos
could be included as sub-grid physics as well, see \citet{MH_sim}. This 
results in slower propagation of the ionization fronts and further delay 
of the final overlap. The halos that can cool atomicly are assumed here 
to be ionizing sources and their recombinations are thus implicitly included 
in the photon production efficiency $f_\gamma$ through the corresponding 
escape fraction.

\section{kSZ from Patchy Reionization}
\label{kSZ_sect}

The kSZ effect is the CMB temperature anisotropy along a line-of-sight
(LOS) defined by a unit vector ${\bf n}$ induced by Thomson scattering
from flowing electrons: 
\be \frac{\Delta T}{T_{\rm CMB}} =\int
d\eta e^{-\tau(\eta)}an_e\sigma_T{\bf n} \cdot {\bf v},
\label{ksz_int}
\ee
where $\eta=\int_0^t dt'/a(t')$ is the conformal time, $a$ is the scale
factor, $\sigma_T=6.65\times10^{-25}\,\rm cm^{-2}$ is the Thomson scattering
cross-section, and $\tau$ is the corresponding optical depth.

We calculate the kSZ anisotropy signal from our simulation data as follows. 
We first calculate the line-of-sight integral in equation~(\ref{ksz_int}) for 
each individual output time-slice of the radiative transfer simulation and for 
all LOS along each of the three box axes. The contribution to the total LOS 
integral from each light-crossing time of the box is then obtained by linear 
interpolation between the nearest results from the simulation output times. 
The individual light-crossing time contributions are then all added together 
to obtain the full LOS integral given in equation~(\ref{ksz_int}). All these 
integrals are done for each LOS through the box along the direction of light 
propagation, which allows us to produce kSZ maps, in addition to the 
statistical signals. In order to avoid artificial amplification of the
fluctuations resulting from repeating the same structures along the 
line of sight, after each light-crossing time we randomly shift the box 
in the directions perpendicular to the LOS and by rotating the box, so that 
the LOS cycles x, y and z axes of the simulation volume.

In addition to our simulations, for comparison we also consider two
simplified cases, of instant and of uniform reionization. We define
instant reionization as a sharp transition from completely neutral to
fully-ionized IGM at redshift $z_{\rm instant}$. We pick
$z_{\rm instant}=13$, which yields the same integrated electron
scattering optical depth as our simulation f250. We also define a
``uniform reionization'' scenario to be one that has the same
time-dependent reionization history (and hence, also the same
$\tau_{\rm es}$) as simulation f250, but spatially uniform (i.e. not
patchy). We then use the same density and velocity data and same
procedures as for the actual simulations to derive the kSZ temperature
fluctuations for these two scenarios. We consider these simple models
in order to demonstrate the effects of reionization being extended and
patchy in nature.

\begin{figure*}[!ht]
\includegraphics[width=3in]{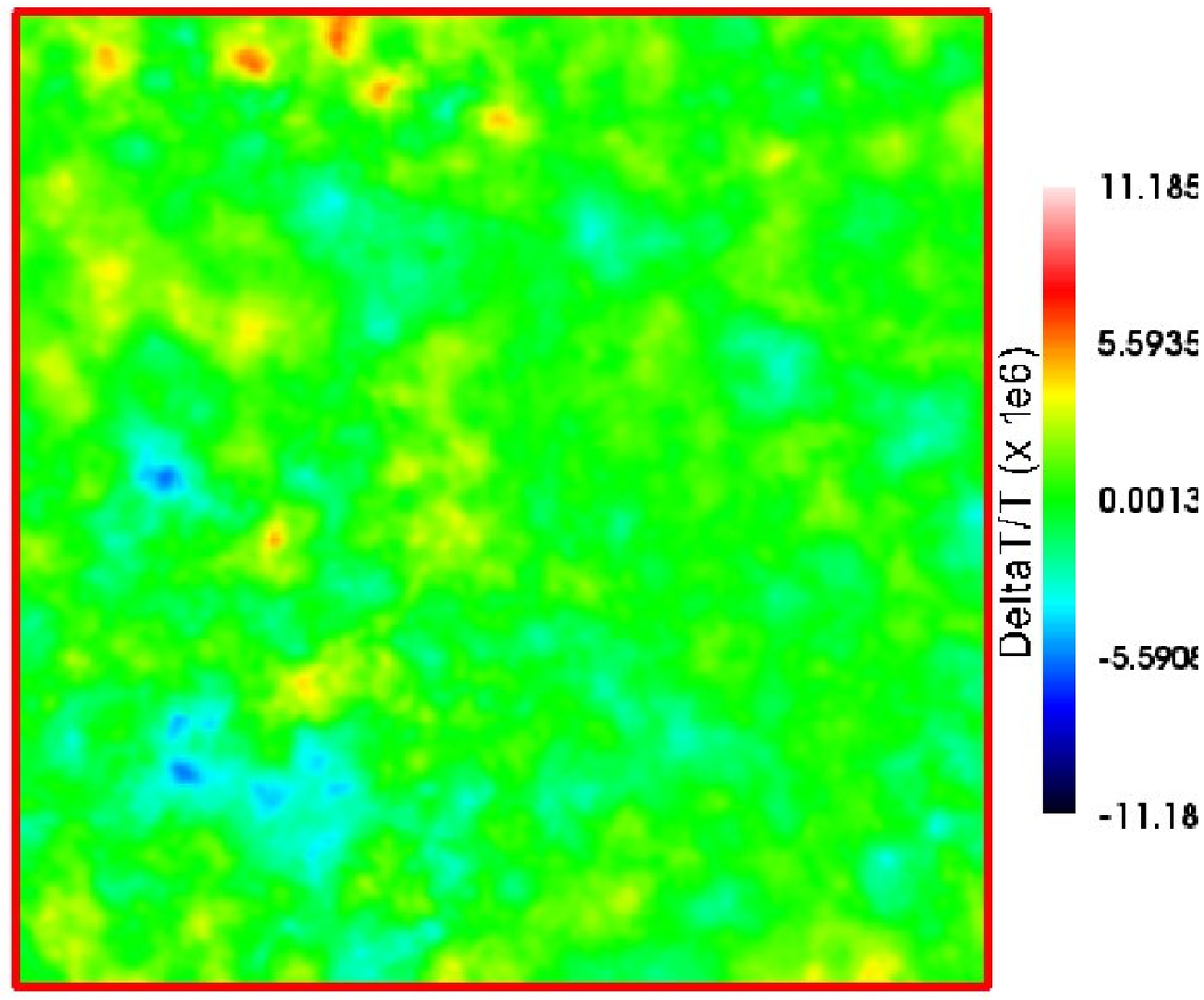}
\includegraphics[width=3in]{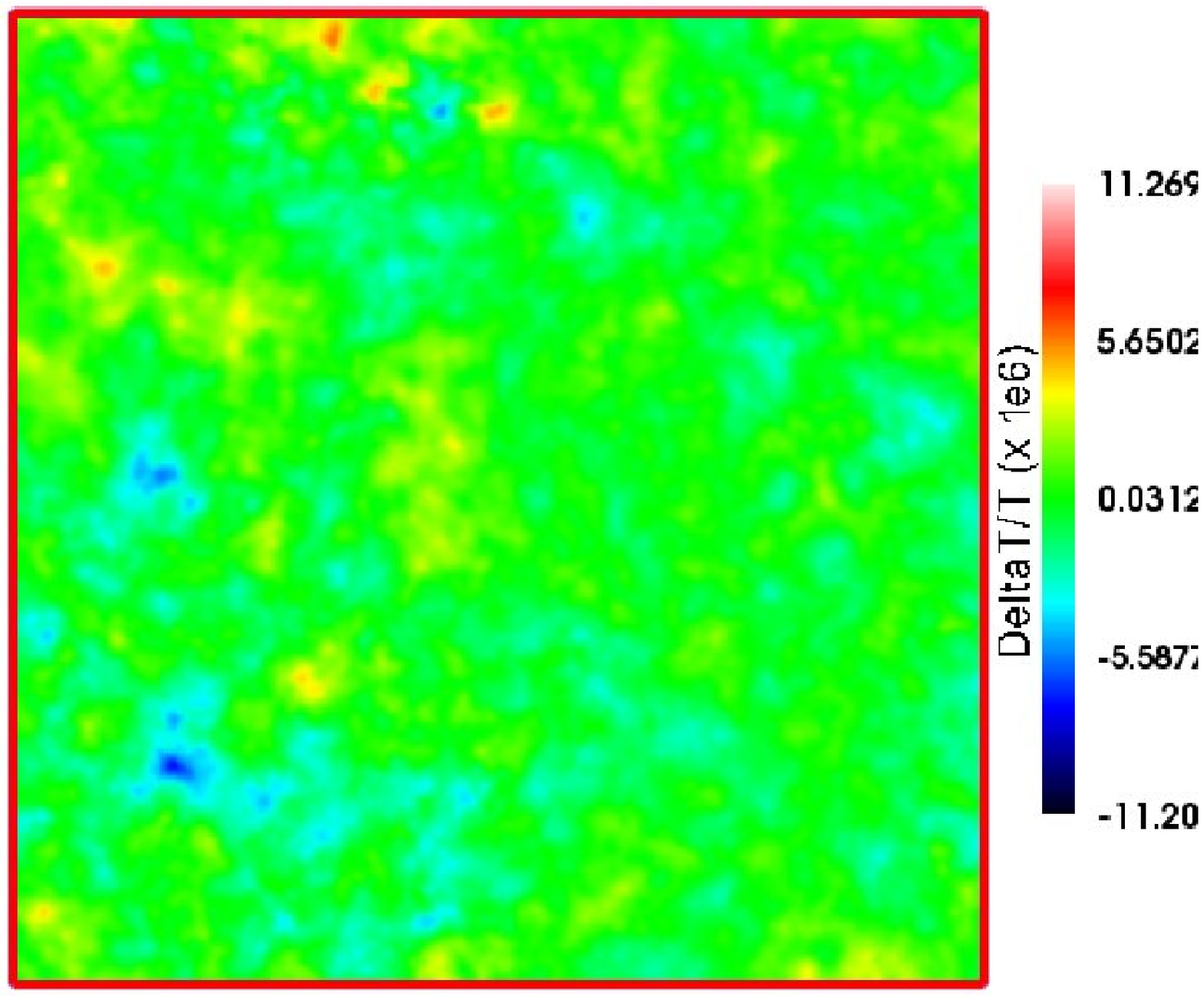}
\caption{\label{maps1} kSZ maps from simulations f2000 (left) and f250
  (right). (Images produced using the Ifrit visualization package of 
  N. Gnedin). 
}
\end{figure*}

\begin{figure*}[!ht]
\includegraphics[width=3in]{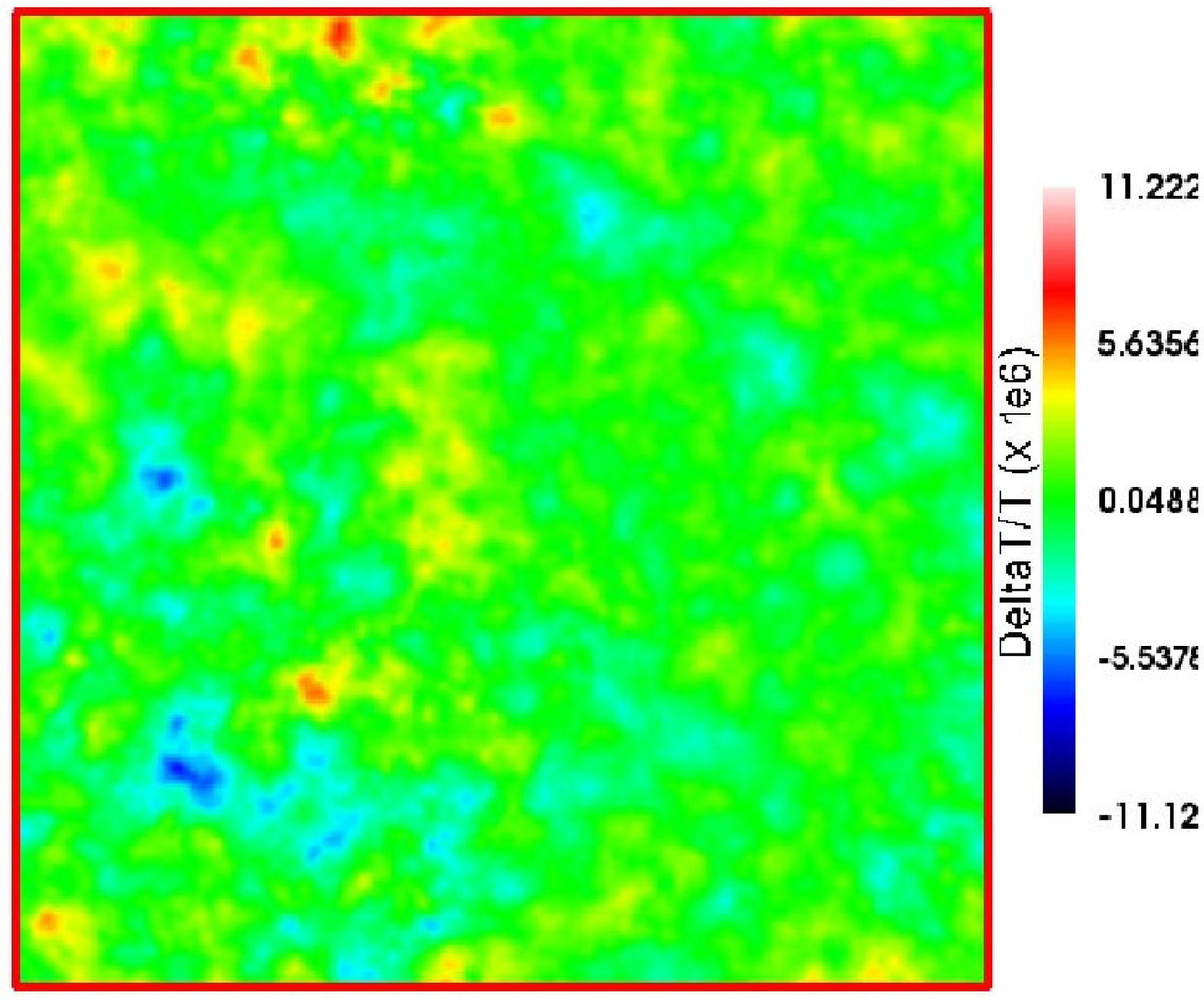}
\includegraphics[width=3in]{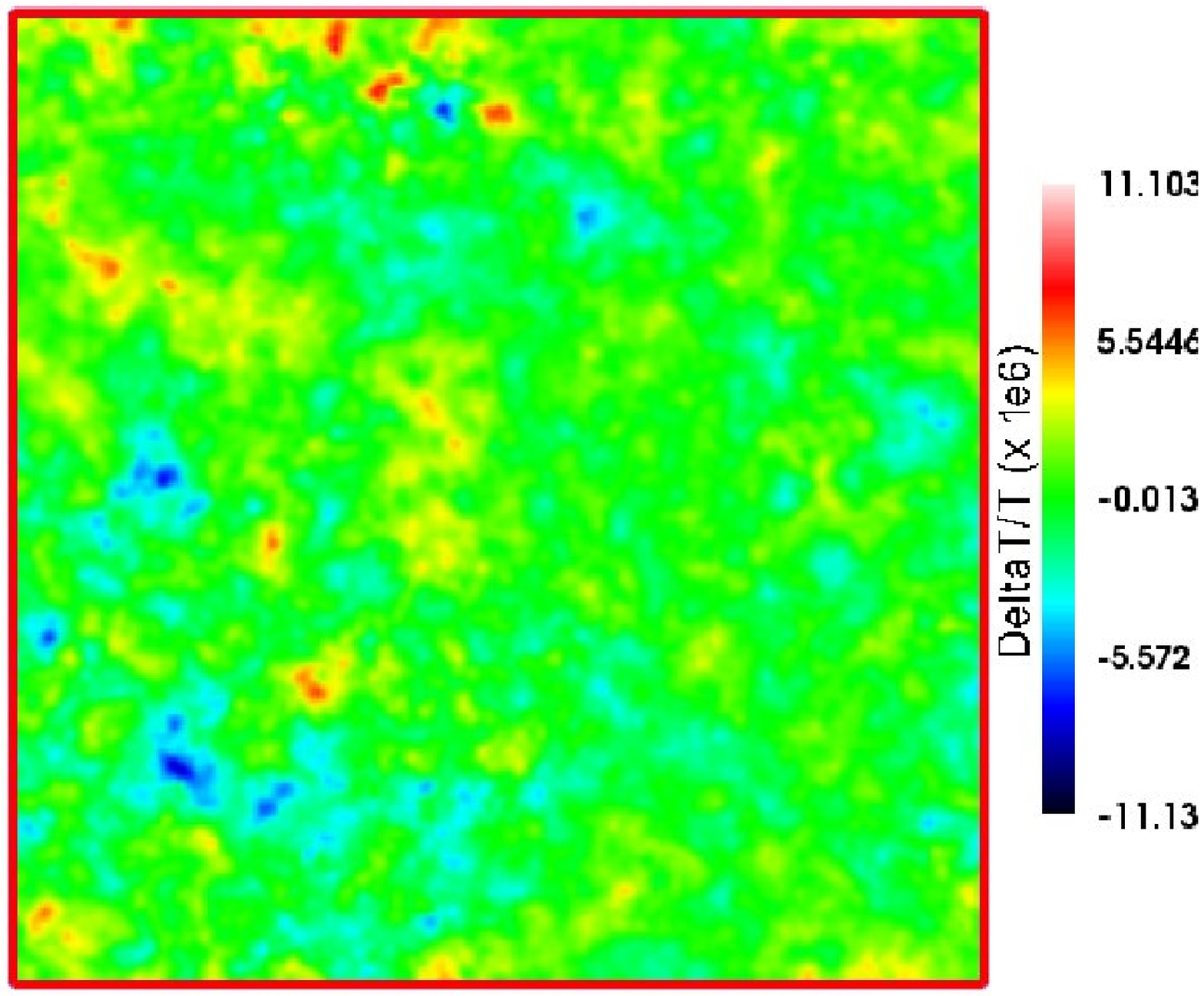}
\caption{\label{maps2} kSZ maps from simulations f2000C (left) and f250C
  (right). 
}
\end{figure*}

\begin{figure*}[!ht]
\includegraphics[width=3in]{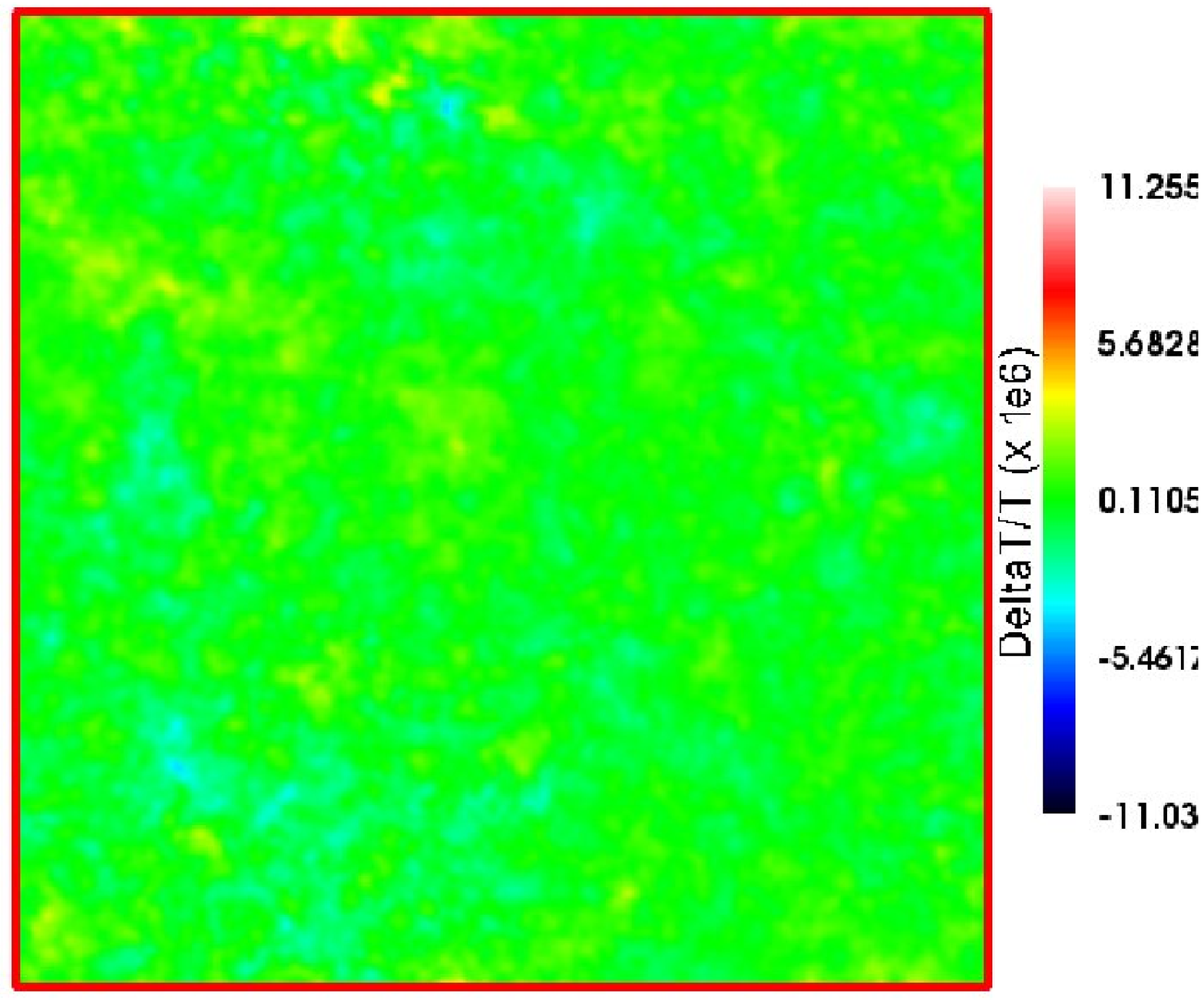}
\includegraphics[width=3in]{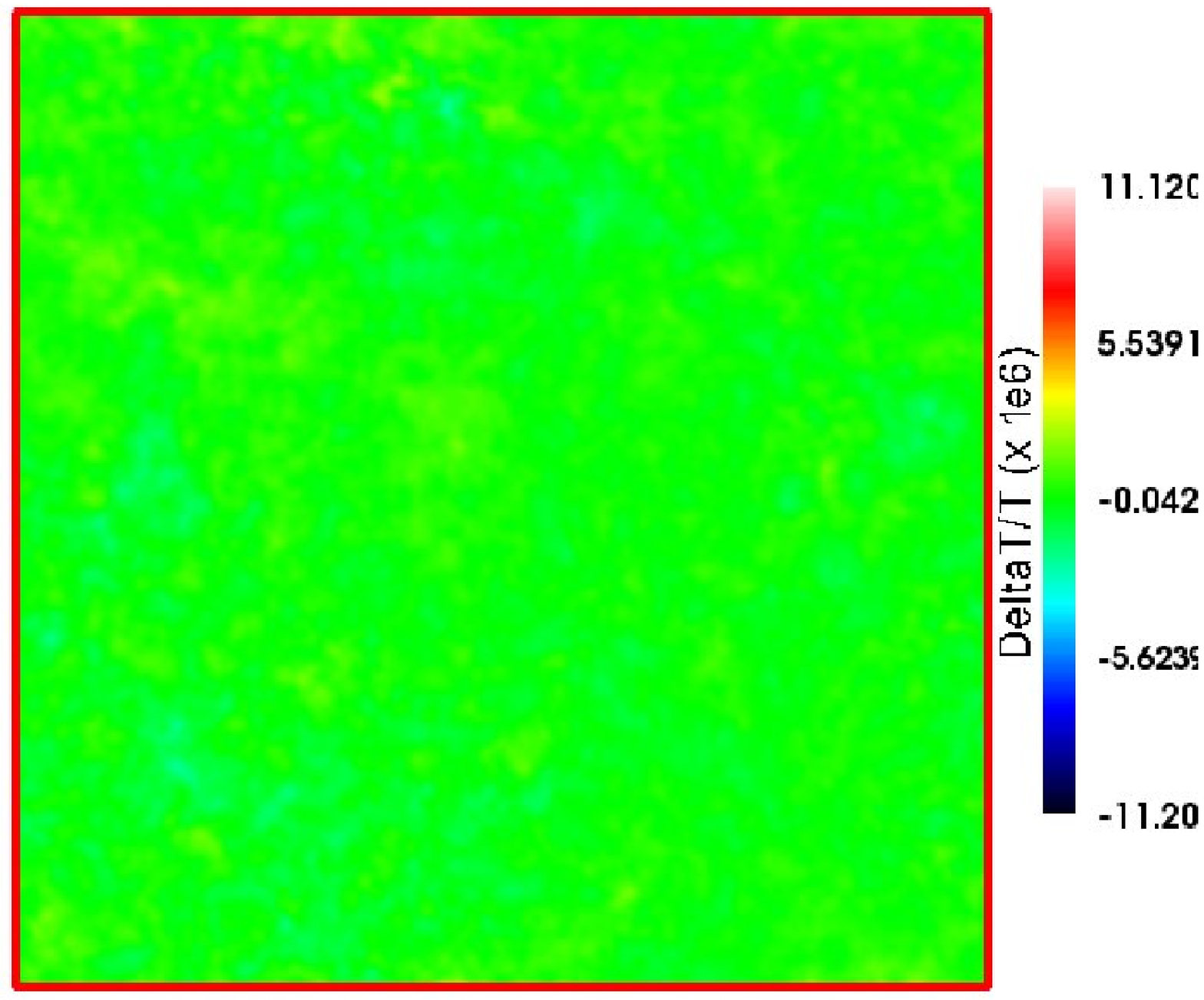}
\caption{\label{maps3} kSZ maps from simulations: (left) assuming 
  instant reionization at $z_{\rm instant}=13$ which gives the same 
  total electron scattering optical depth as simulation f250, and 
  (right) spatially-uniform reionization with the same reionization 
  history and thus same total electron scattering optical depth as 
  simulation f250.
}
\end{figure*}

\begin{table*}
\caption{Mean and rms values for $\delta T_{\rm kSZ}/T_{\rm CMB}$.}
\label{rms}
\begin{center}
\begin{tabular}{@{}lllllll}
%
simulation & f2000  &                 f2000C    &             f250        &             f250C & uniform & instant\\[2mm]\hline
mean    & $8.86\times10^{-8}$ & $9.88\times10^{-8}$ & $-2.17\times10^{-7}$&$5.57\times10^{-8}$&$-3.29\times10^{-12}$&$-3.37\times10^{-12}$\\[2mm] 
rms     & $1.13\times10^{-6}$ & $1.22\times10^{-6}$ & $1.06\times10^{-6}$ &$1.09\times10^{-6}$&$4.24\times10^{-7}$&$6.67\times10^{-7}$\\[2mm]
\hline\\
\end{tabular}
\end{center}
\end{table*}

\section{Results}

We show the kSZ maps of temperature fluctuations, $\delta T/T_{\rm
CMB}$, yielded by all of our cases in Figures~\ref{maps1} (f2000 and
f250), \ref{maps2} (f2000C and f250C), and \ref{maps3} (instant and
uniform reionization). These have a total angular size of
approximately $50'\times50'$, corresponding to our computational box
size. The resolution of the maps is that of the full radiative
transfer grid ($203\times203$ pixels), corresponding to pixel
resolution of $\sim0.25'$. All maps utilize the same color map in
order to facilitate their direct comparison. The maps deriving from
our simulations of inhomogeneous reionization (Figures~\ref{maps1} and
\ref{maps2}) all show fairly strong fluctuations, both positive and
negative, of order $\delta T\sim10\, \mu K$ at angular scales of a few
arcminutes ($\rm\sim10~h^{-1}Mpc$) and up to $\delta T\sim20\, \mu K$
at the full map resolution. The fluctuations are at somewhat smaller
scales when the sources are less efficient photon producers (f250 and
f250C), compared to the high-efficiency cases (f2000 and f2000C). The
variations are noticeably enhanced when sub-grid gas clumping is
included in the reionization model (f2000C and f250C) and there are a
number of regions with very strong features. In comparison, the
artificial models of instant and uniform reionization
(Figure~\ref{maps3}) show fluctuations with much lower amplitudes and
with a less well-defined typical scale. Since these simple scenarios
were constructed to produce the same total electron scattering optical
depth as simulation f250 and they use the same density and velocity
fields as the simulations, any differences we see are due to the
patchiness of reionization.  The lower signal in these last two cases
is expected since the kSZ effect in uniformly-ionized gas exactly
cancels in the first order. This cancellation is broken when
patchiness is present, enhancing the anisotropies.

The mean and rms of the temperature fluctuations for all cases are summarized
in Table~\ref{rms}. Based on these, we see that, indeed, the uniform and
instant reionization cases have means very close to zero, while the realistic, 
patchy reionization simulations yield mean temperature fluctuations that are
low, of order $\sim10^{-7}$, but not zero. The rms values are 
$\sim10^{-6}$ for all patchy reionization cases and lower than that by factor 
of $\sim2$ ($\sim3$) for the instant (uniform) reionization cases.


\begin{figure}
\includegraphics[width=3.7in]{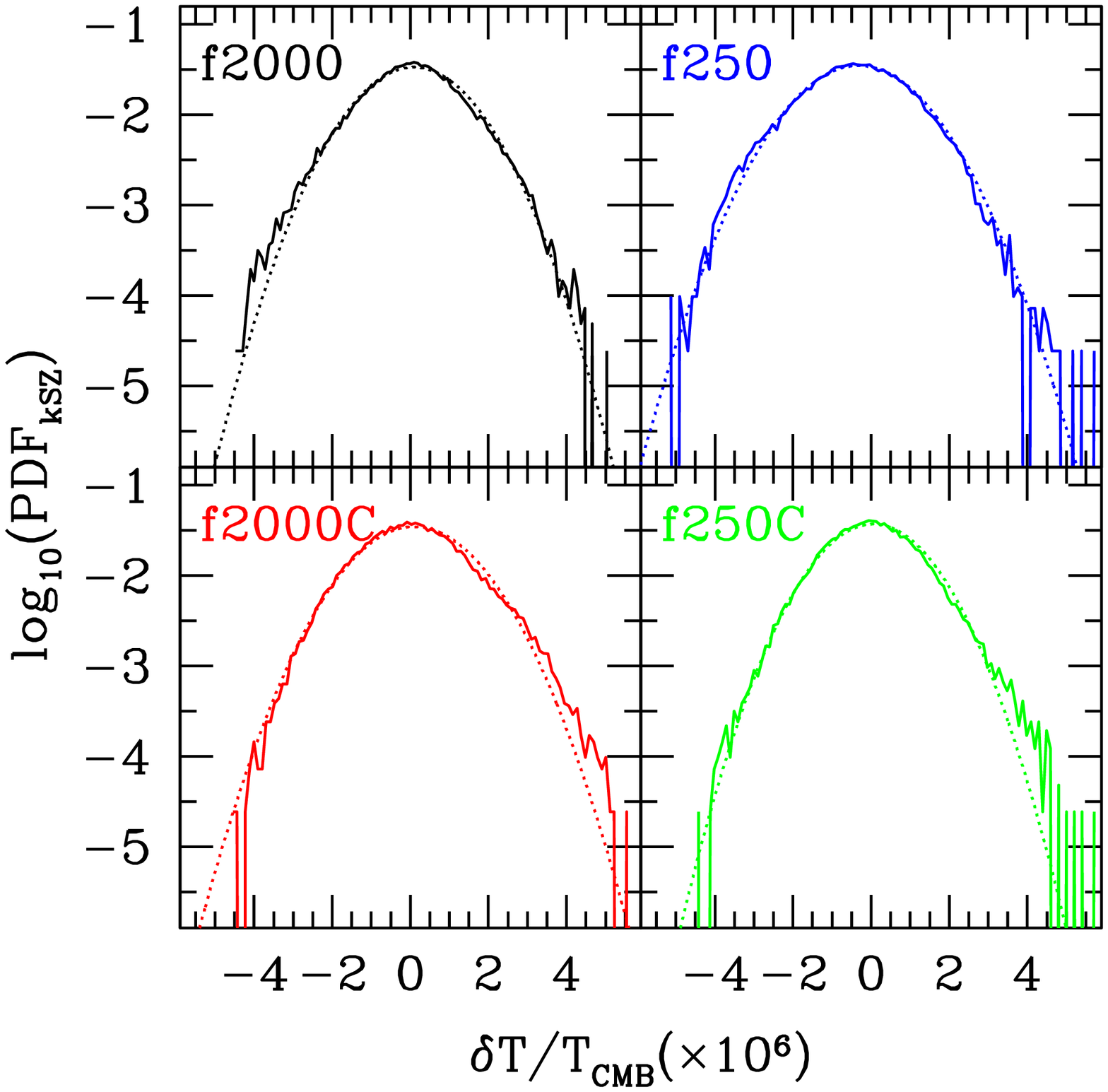}
\caption{\label{PDF_fig} PDF distribution of $\delta T_{\rm kSZ}/T_{\rm CMB}$ 
(solid) vs. Gaussian distribution with the same mean and width (dotted).  
}
\end{figure}

\begin{figure}[!ht]
\includegraphics[width=3.7in]{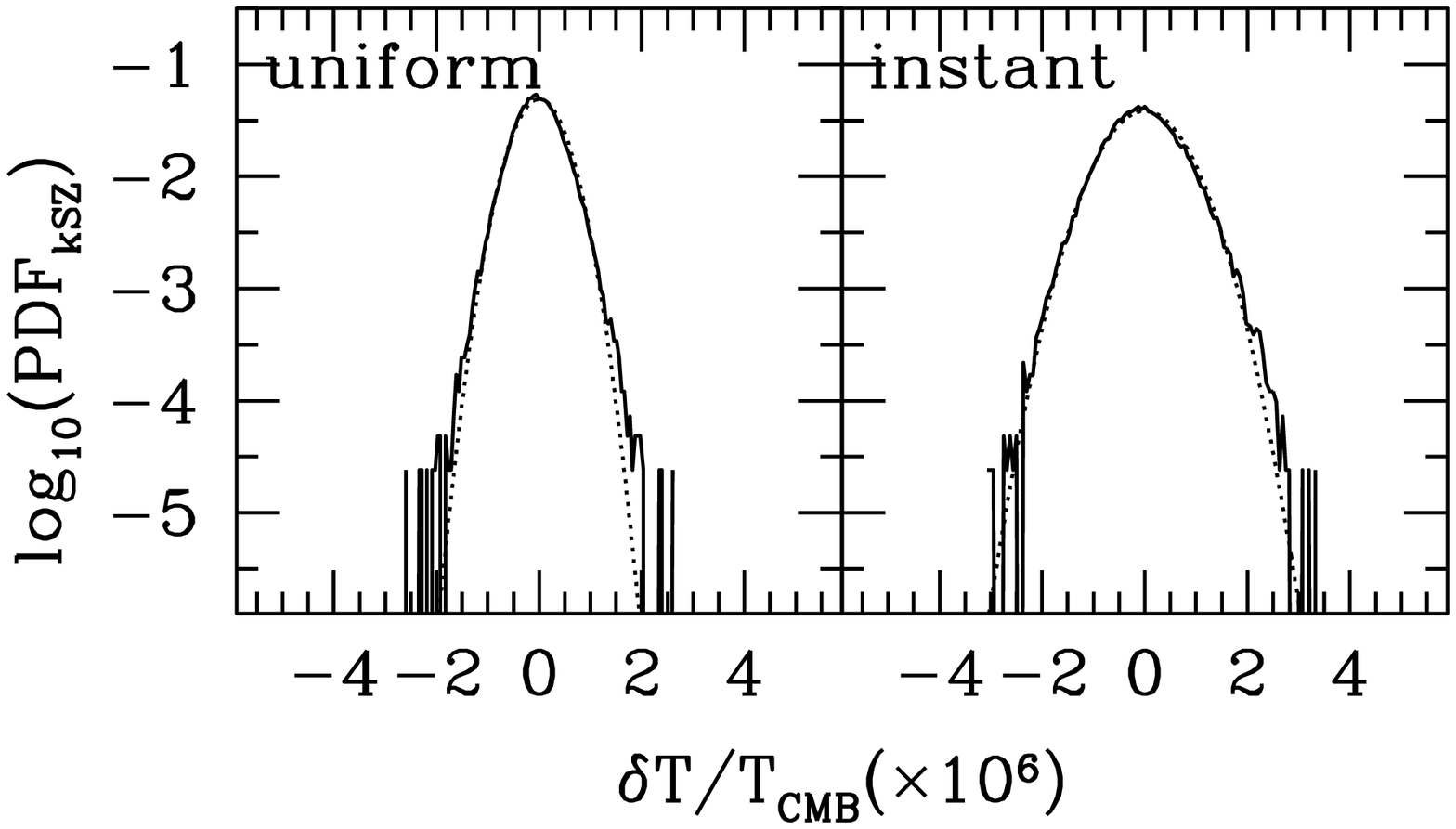}
 \vspace{-1.5in} 
\caption{\label{PDF_special_fig} PDF distribution of 
$\delta T_{\rm kSZ}/T_{\rm CMB}$ (solid) vs. Gaussian distribution with 
the same mean and width (dotted).  
}
\end{figure}

These observations based on the maps are confirmed by their
corresponding pixel PDF distributions, shown in Figures~\ref{PDF_fig}
and \ref{PDF_special_fig}. On each panel we also plot the Gaussian
distribution with the same mean and standard deviation. All
distributions are surprisingly close to Gaussian around their mean
values given the maps. However noticeable departures from Gaussianity
do occur in the wings of the PDFs. In particular, the PDFs for patchy
reionization with sub-grid clumping (f2000C, and f250C) are
significantly non-Gaussian. For the realization simulated there is an
over-abundance of bright regions by up to an order of magnitude
compared to the corresponding Gaussian distributions. The reionization
scenarios without sub-grid clumping show much weaker non-Gaussian
features there. This indicates that the observed PDF of kSZ from
patchy reionization may give us important information on the level of
small-scale gas clumpiness during reionization.  The PDFs derived from
the uniform and instant reionization scenarios are significantly less
wide than the simulated ones (which was also shown by their lower rms
values, as was discussed above), and are also much closer to Gaussian.

\begin{figure}[!ht]
\includegraphics[width=3.7in]{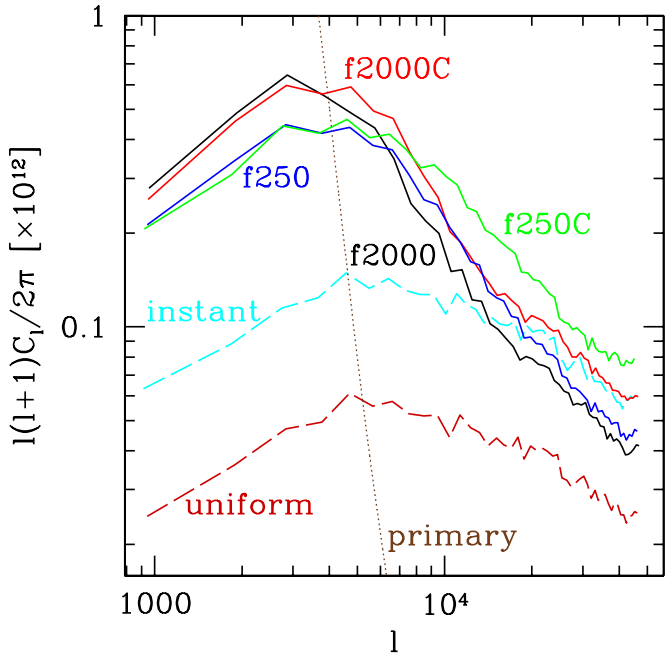}
\caption{\label{ps} Sky power spectra of $\delta T_{\rm kSZ}/T_{\rm CMB}$
  fluctuations resulting from our simulations: f2000 (black), f250
  (blue), f2000C (red) and f250C (green). For comparison, we also show the 
 results from simple models which utilize the same density and velocity fields 
 as the actual simulations, but different assumptions about the gas ionization 
 in space and time, a uniform reionization with the same reionization history
  as simulation f250 (dark red, short-dashed) and an instant reionization
  model with the same integrated optical depth $\tau_{\rm es}$ as simulation 
 f250 (cyan, short-dashed). The primary CMB anisotropy signal is also shown 
 (brown, dotted).} 
\end{figure}

The kSZ anisotropy signal from inhomogeneous reionization dominates
the primary CMB anisotropy above $\ell=3000$. The sky power spectra
(Figure~\ref{ps}) peak strongly at $\ell=3000-5000$, to a maximum of
$[\ell(\ell+1)C_\ell/(2\pi)]_{\rm max}\sim7\times10^{-13}$ when the
ionizing sources are highly-efficient (cases f2000 and f2000C). The
peak values for the simulations with lower-efficiency sources (f250
and f250C) are only slightly lower, at
$[\ell(\ell+1)C_\ell/(2\pi)]_{\rm max}\sim4\times10^{-13}$, and the
peaks are somewhat broader and moved to smaller scales,
$\ell\sim3000-7000$, than for the high-efficiency cases. The
introduction of sub-grid gas clumping in the radiative transfer
simulations produces significantly more power at small scales and
slightly broader peaks when compared to the similar cases without
sub-grid clumping. The scales at which the power spectra peak roughly
correspond to the typical ionized bubble sizes observed in our
simulations, namely $\sim5-20$~Mpc comoving. These typical sizes
depend on the assumed source efficiencies and sub-grid clumping. At
large scales the patchy kSZ signal depends only on the source
efficiencies, being higher for more efficient photon emitters, since
these tend to produce larger ionized regions on average.

\begin{figure*}
\includegraphics[width=2.3in]{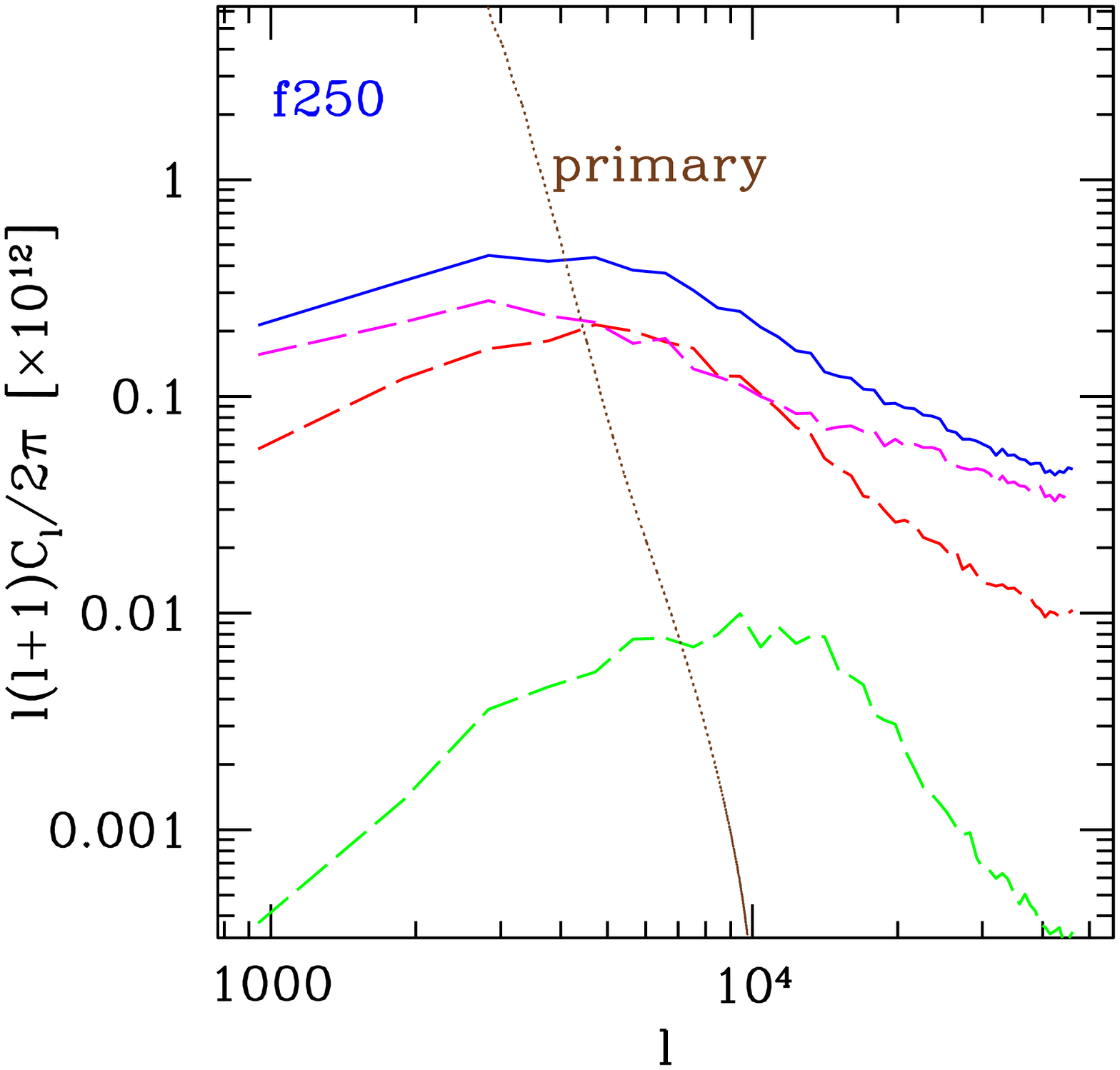}
\includegraphics[width=2.3in]{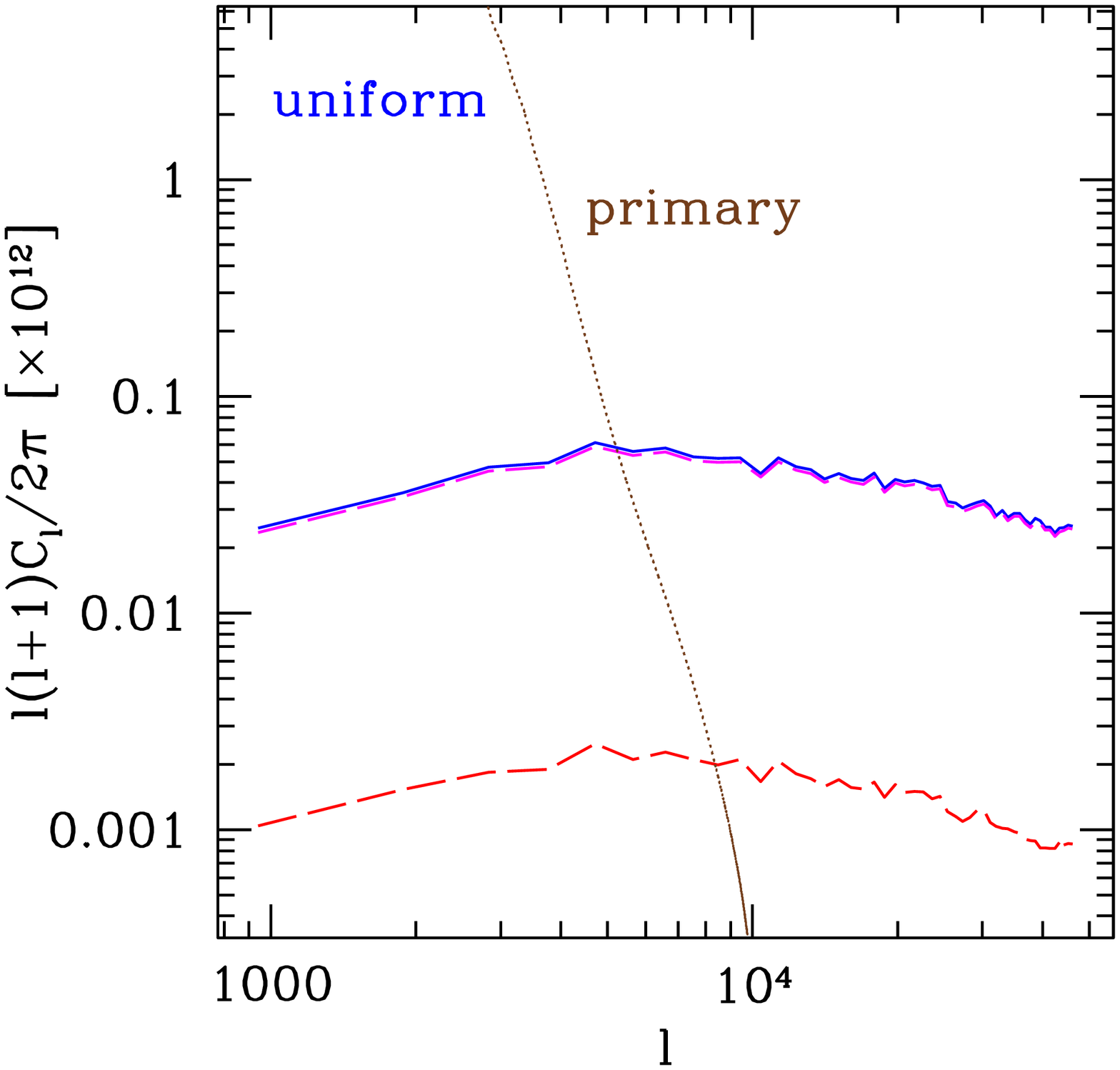}
\includegraphics[width=2.3in]{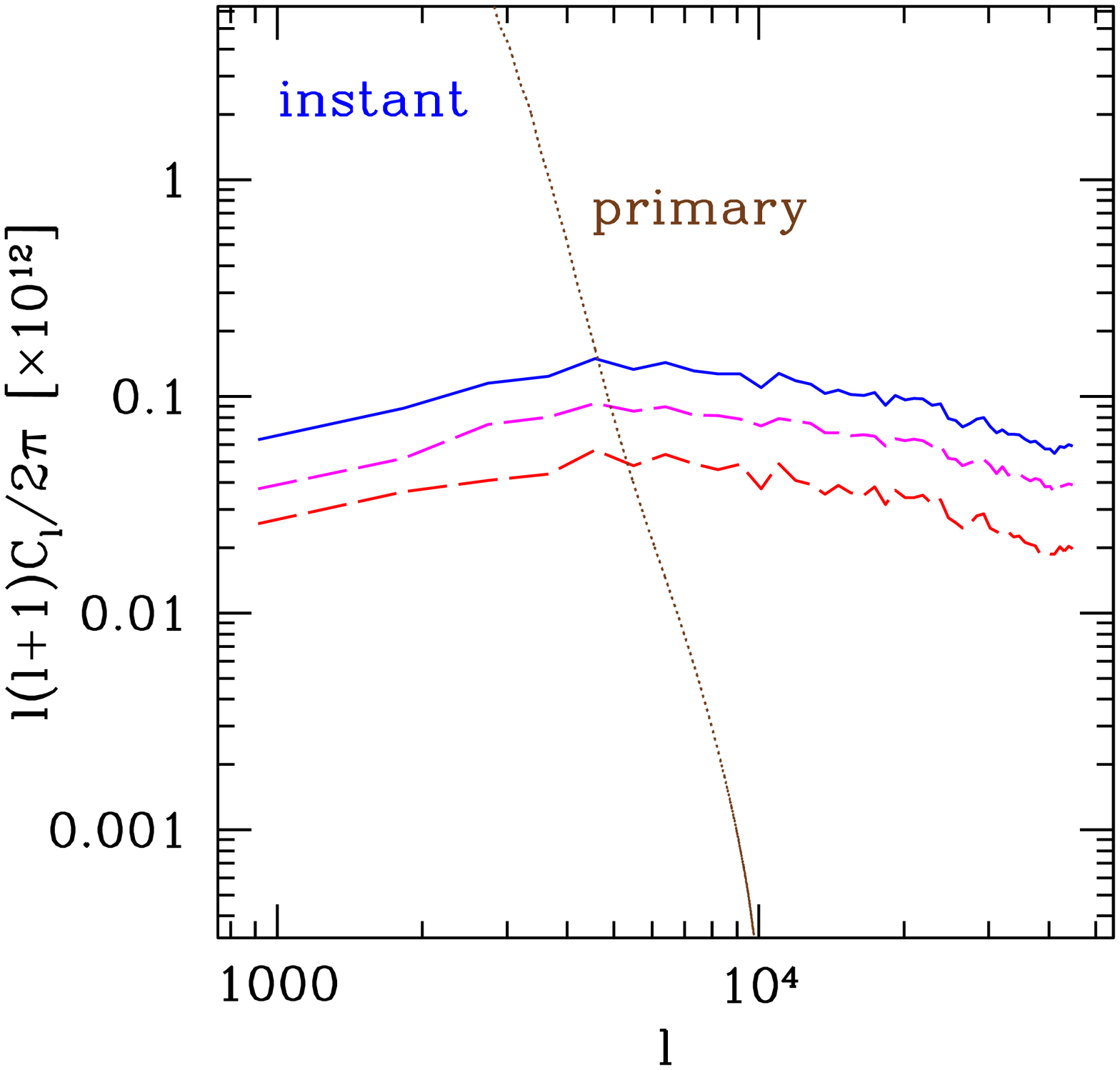}
\caption{\label{fig4} Contributions to the total kSZ signal (blue) from 
different redshifts for case f250 (left), uniform (middle) and instant 
reionization (right). Shown are the signals for every three box light 
crossing times, roughly corresponding to (bottom to top on the left) 
redshifts $z>15$ (green), $15>z>11$ (red), $z<11$ (magenta). In the last 
two cases the $z>15$ contribution is very low (uniform) or zero (instant) 
and thus not show.}
\end{figure*}

In contrast, the uniform reionization scenario (with the mean reionization
history of simulation f250) yields a kSZ signature which is much lower than 
the non-uniform reionization scenarios, indicating a very large boost of the
signal due to the patchiness. The boost is largest, approximately one order of
magnitude, at and above the typical scale of the patches ($\ell<7000$), but is
still exists at smaller scales, where it is a factor of two or more. The other
simplified  scenario, of instant reionization with the same integrated optical
depth at f250, produces larger kSZ anisotropy than a uniform reionization
does. However, it is still well below the realistic patchy reionization
signals, by factor of $\sim3$ for $\ell<7000$. At smaller scales the
reionization 
signal from the instant reionization scenario becomes similar to the ones from 
simulated non-uniform reionization. The distribution of the power in these two 
simplified scenarios is much flatter, with less indication of a characteristic 
scale. This implies that the sharp peaks yielded by the inhomogeneous
reionization scenarios are dictated by the size of the ionized patches, since
both the density and the velocity fields are shared among the models.

This behaviour can be understood further by considering the contributions from
different redshift intervals to the integrated signal. In Figure~\ref{fig4} we 
show the contributions to the total signal from different redshift intervals  
for simulation f250, as well as the uniform and instant reionization scenarios.
We plot the contribution from every three light-crossing times of the box, 
corresponding roughly to $z>15$, $15>z>11$ and $z<11$. The mass-weighted 
ionized fraction in the highest redshift interval is $x_m<0.01$, and
consequently its contribution to the kSZ effect (case f250, bottom curve) is
low, with a maximum of $\sim10^{-14}$. This high-redshift contribution is
strongly peaked at very small scales ($\ell>10^4$), reflecting the fact that
at that time the typical ionized bubble is still quite small, less than few
Mpc in size. The contribution from the middle redshift interval, $11<z<15$,
(corresponding to $0.8>x_m>0.01$) peaks at roughly the same scales as the
integrated signal. It contributes about half of the total signal around the
peak, but a smaller fraction at scales above and below that. A half or more of
the integrated kSZ signal at any scale is contributed by the lowest redshifts
($z<11$, $x_m\gtrsim0.8$). The low-redshift power spectrum is fairly flat, 
with a weak peak at relatively large scales, $\ell\sim2000$. 

In the homogeneous (uniform and instant) reionization cases the shape of the
power spectra is essentially the same for all redshift intervals, quite flat, 
with a broad peak at $\ell\sim5000$. The signal from the
uniform reionization scenario is completely dominated by the lowest redshifts,
with less than a few per cent coming from the middle redshift interval, and
essentially no contribution from high redshifts. Since the velocity and
density fields are shared, the instant reionization scenario contributions
differ from the uniform scenario ones only by being weighted by the ionized
fraction in the uniform case. The instant reionization scenario assumes full
ionization after $z_{\rm instant}=13$ and fully-neutral gas before that. Thus,
as a consequence to the kSZ signal being stronger at later times, the instant
reionization integrated signal is higher, as well. 

\begin{figure}[!ht]
\includegraphics[width=3.7in]{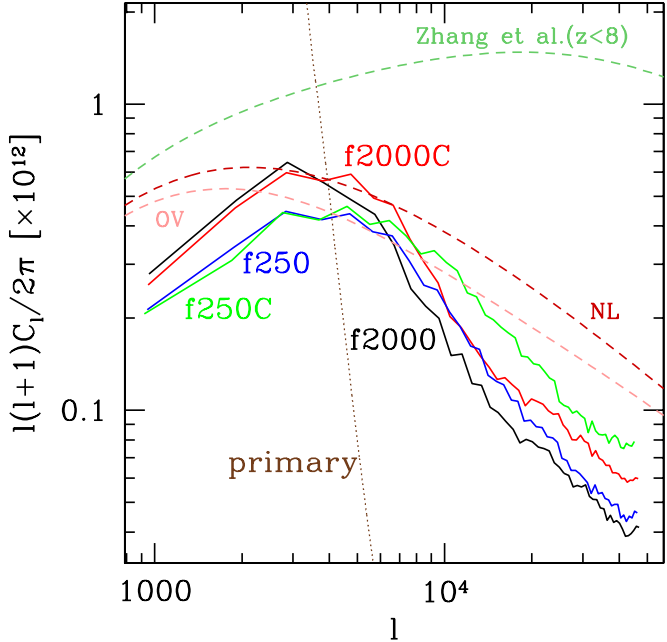}
\caption{\label{ps_after} Sky power spectra of $\delta T_{\rm
  kSZ}/T_{\rm CMB}$ fluctuations from our simulations f2000 (black),
  f250 (blue), f2000C (red) and f250C (green) compared to the
  after-reionization kSZ signals (assuming overlap at $z_{\rm ov}=8$):
  linear Ostriker-Vishniac effect, labelled 'OV' (long-dashed, pink),
  the same, but using the nonlinear power spectrum of density
  fluctuations, labelled 'NL' (long-dashed, dark red) and a
  fully-nonlinear model matched to high-resolution hydrodynamic
  simulations of \citet{2004MNRAS.347.1224Z}, labelled `Zhang et al'
  (short-dashed, dark green).  The primary CMB anisotropy signal is
  also shown (brown, dotted).}
\end{figure}

Finally, we compare the two contributions to the total kSZ
anisotropies, the patchy reionization signal we have calculated and
the kSZ signal from fully-ionized gas after reionization. In
Figure~\ref{ps_after} we show our predictions against three
representative calculations of the post-reionization contribution: the
quadratic-order Ostriker-Vishniac effect \citep{1987ApJ...322..597V}
expressed in terms of a product of the linear power spectrum; the same
expression, but with the nonlinear density power spectra substituted
for one of the linear ones, which partially accounts for the nonlinear 
effects; and the recent detailed nonlinear model of
\citet{2004MNRAS.347.1224Z}. The last three models are rescaled to our
adopted cosmology using $\ell(\ell+1)C_\ell\propto \sigma_8^5$
\citep{2004MNRAS.347.1224Z}, and assume reionization overlap at
$z_{\rm ov}=8$, to allow for direct comparison. Confronting first the
three uniform-ionization models, we see that they agree fairly well at
large, linear scales, but strongly diverge, by up to one order of
magnitude at small scales, where the nonlinearities become very
important. Including the corrections due to the nonlinear density
power spectrum yields modestly higher kSZ signal than the linear OV
calculation, and similar power spectrum shape overall. Both peak at
roughly the same scale, $\ell\sim2000$. In contrast, the
\citep{2004MNRAS.347.1224Z} result finds much larger signal,
especially at small scales, which peaks at $\ell\sim20,000$.  Compared
to our patchy reionization kSZ predictions, all uniform-ionization
power spectra are much less sharply peaked, i.e. they lack a
well-established characteristic scale. In terms of their peak values,
the OV and OV with nonlinear corrections models yield values similar
to our patchy reionization results, while \citep{2004MNRAS.347.1224Z}
find almost a factor of 2 higher signal at the patchy signal peak, and
even more pronounced dominance at smaller scales.

\section{Summary and Discussion}

We have derived the kSZ CMB anisotropies due to the inhomogeneous reionization
of the universe. This is the first such calculation based on detailed,
large-scale radiative transfer simulations of this epoch. The resulting sky
power spectra peak at $\ell=2000-8000$ with maximum values of 
$[\ell(\ell+1)C_\ell/(2\pi)]_{\rm max}\sim4-7\times10^{-13}$. This roughly
corresponds to the typical ionized bubble sizes observed in our simulations, 
which is $\sim5-20$~Mpc, depending on the assumed source efficiencies and
the gas clumping at very small scales. The kSZ anisotropy signal from
reionization dominates the primary CMB signal above $\ell=3000$. At large 
scales the patchy kSZ signal depends only on the source efficiencies. It is
higher when sources are more efficient at producing ionizing photons, since
such sources produce large ionized regions, on average, than less efficient
sources. The introduction of sub-grid gas clumping in the radiative transfer 
simulations produce significantly more power at small scales, but has little
effect at large scales. The integrated kSZ signal is strong enough to be
detected by upcoming experiments, like ACT and SPT, but it seems difficult
to separate the patchy reionization signal from the contribution by the
fully-ionized gas after reionization.

Previous simulations of the kSZ effect from patchy reionization 
\citep{2001ApJ...551....3G,2005MNRAS.360.1063S} predicted significantly lower
kSZ signals than the ones we found. Their power spectra reach maximum values
of $[\ell(\ell+1)C_\ell/(2\pi)]_{\rm max}\sim2\times10^{-14}$ and 
$\sim1.6\times10^{-13}$, respectively, compared to 
$[\ell(\ell+1)C_\ell/(2\pi)]_{\rm max}\sim4-7\times10^{-13}$ for our
simulations. This discrepancy is due to the small volumes used in these
simulations, which significantly reduces the power in both the density and  
the velocity field fluctuations. At the scale of the simulation box the bulk 
velocities are zero by definition, and the density is the mean one for the 
universe, thus any larger-scale fluctuations are not included, and the ones 
close to the box size are underestimated. In terms of the calculation of kSZ
anisotropy from patchy reionization, this problem is more severe for the
velocity field, since its power grows towards large scales, unlike the density
field fluctuations, which instead decrease at large scales. The reionization
patchiness imprints its characteristic scale on the density and velocity
fluctuations, effectively smoothing the small-scale fluctuations below the
typical bubble sizes. On the other hand, the large-scale fluctuations still
should contribute to the kSZ signal, since the H~II regions are moving with
the large scale bulk motions. 

In addition to the missing large-scale power, these earlier reionization 
simulations also did not follow sufficient volumes to properly sample the 
size distribution of the ionized bubbles. The typical sizes of these ionized 
patches are of order 5-20 comoving Mpc, and even larger at later times, and 
hence require simulation volumes of order $(100\,h^{-1}Mpc)^3$ for correct 
sampling. As a result, the kSZ power spectra found by the smaller-box 
simulations were largely flat, with no clear characteristic scale, since they
do not reach the large characteristic bubble scales, which are dictated by the
strong clustering of ionizing sources at high redshifts.  

Several semi-analytical models for calculating the kSZ signature of
patchy reionization have been proposed in recent years
\citep{1998ApJ...508..435G,
2003ApJ...598..756S,2005ApJ...630..643M}. \citet{1998ApJ...508..435G}
proposed a very simple model, whereby the ionized patches are randomly
distributed and have a given characteristic size $R$. Furthermore,
they assumed that the density, velocity and ionization fraction fields
are all uncorrelated with each other.  The ionized fraction
auto-correlation function is approximated as a Gaussian with rms given
by the (fixed) characteristic H~II region size. Under these
assumptions the kSZ power spectrum can be calculated analytically. The
quantitative details of the signal predicted by this model depend on
the assumptions made about the typical ionized patch size, and the
time and duration of reionization, but generically it predicts a
distribution which is very strongly peaked, more so than our
simulations find. This is related to the assumed Gaussian distribution
around a fixed characteristic size of the ionized patches. The actual
simulations also find a characteristic size for the ionized bubbles,
but one that is evolving with redshift, and different size
distributions around it, resulting in somewhat less sharp peaks than
this analytical model predicts.
  
\citet{2003ApJ...598..756S} proposed another simple model for the
reionization patchiness, which assumes that the correlation function
of the ionized patches is proportional to the one for the density
field, with a bias factor which is time-dependent, but not
scale-dependent. The power is filtered at small scales, below the
typical size of the ionized bubbles using a Gaussian filer in k-space.
The time-dependence of the typical size of the ionized patches is
assumed to be proportional to $(1-x_m(t))^{-1/3}$, where $x_m(t)$ is
the mean mass-weighted ionized fraction at time $t$. The last function
is based on the semi-analytical models of reionization of
\citet{2003ApJ...595....1H}. The resulting kSZ power spectra are
fairly flat, with much less well-defined characteristic scale than our
simulation results, as should be expected based on the
quickly-evolving typical patch size assumed in this model. The peak
amplitudes of the results is somewhat higher than the ones we find, by
factors of $\sim2-3$, probably due to their assumed values of the
effective bias.

More recently, \citet{2005ApJ...630..643M} applied the 
semi-analytical reionization model of \citet{2004ApJ...613....1F} to
estimate the kSZ signal.  They found lower fluctuations in their
``single reionization episode'' models (by factor of $\sim2$),
although they have similar mean reionization histories (i.e. evolution
of the mean ionized fraction) to our
simulations. Their results predict a peak at somewhat larger scales
($\ell\sim$2000) than ours ($\ell\sim$2000-9000). Their ``extended
reionization'' scenarios peak higher (in some cases slightly above
$[\ell(\ell+1)C_\ell/(2\pi)]_{\rm max}\sim10^{-12}$) again at
$\ell\sim$2000.  We note that these extended scenarios have a very
different assumed reionization histories than any of our
simulations. They found that the kSZ post-reionization contribution
dominates the patchy signal at the scales of interest, in rough agreement
with our results.



\section*{Acknowledgments} 
This work was partially supported by NASA Astrophysical Theory Program grants
NAG5-10825 and NNG04GI77G to PRS. 

\bibliographystyle{elsart-harv}\bibliography{../../../refs}





\end{document}